# Electron-pair densities with time-dependent quantum Monte-Carlo

Ivan P. Christov

Physics Department, Sofia University, 1164 Sofia, Bulgaria

Email: ivan.christov@phys.uni-sofia.bg

**Abstract**

In this paper we use sets of de Broglie-Bohm trajectories to describe the quantum correlation effects which take place between the electrons in helium atom due to exchange and Coulomb interactions. A short-range screening of the Coulomb potential is used to modify the repulsion between the same spin electrons in physical space in order to comply with the Pauli's exclusion principle. By calculating the electron-pair density for ortho-helium we found that the shape of the exchange hole can be controlled uniquely by a simple screening parameter. For para-helium the inter-electronic distance, and hence the Coulomb hole, results from the combined action of the Coulomb repulsion and the non-local quantum correlations. In this way a robust and self-interaction-free approach is presented to find both the ground state and the time evolution of non-relativistic quantum systems.

## 1. Introduction

The electronic many-body problem is of key importance for the theoretical treatments of physics and chemistry. A typical manifestation of the quantum many-body effects is the electron correlation which results from the Coulomb and exchange interactions between the electrons combined with the underlying quantum non-locality. Since in general the electron correlation reshapes the probability density in configuration space, it is difficult to elucidate this effect for higher dimensions. Therefore, to better understand the effects of electron correlation in atoms and molecules one needs, besides one-particle quantities such as the electron density function, to consider also extensions which explicitly incorporate many-body effects. Such an appropriate quantity is the electronic pair density function which represents the probability density of finding two electrons at distance **u** from each other [1]:

$$I(\mathbf{u},t) = \left\langle \Psi(\mathbf{R},t) \left| \sum_{i<j} \delta\left[(\mathbf{r}_i - \mathbf{r}_j) - \mathbf{u}\right] \right| \Psi(\mathbf{R},t) \right\rangle, \tag{1}$$

where $\mathbf{r}_i$ is the position of the $i$th electron and the many-body wave function $\Psi(\mathbf{R},t)$ resides in configuration space with arguments being the instantaneous coordinates of all electrons $\mathbf{R} = (\mathbf{r}_1, \mathbf{r}_2, ..., \mathbf{r}_N)$.

The importance of the electron pair density, also known as electron position intracule, comes from the fact that it can be associated with experimental data obtained from x-ray scattering, and it can also be used to visualize the notion of exchange and correlation holes which surround the quantum particles. However the calculation of the many-body wave function in Eq. (1) is hampered by the computational cost which scales exponentially with system dimensionality. Therefore, different approximations have been employed in order to calculate the electronic pair densities. These include Hartree-Fock (HF) approximation as well as Hylleraas type explicitly correlated wave functions represented as product of HF function and pair-correlation factors [2-5]. Other (e.g. quantum Monte Carlo [6]) approaches use appropriate Slater–Jastrow-type many-body wave functions which involve number of parameters, which after optimization can be used to calculate the average in Eq. (1).

Here we calculate the electron pair densities for helium atom in $2\ ^1S$ and $2\ ^3S$ states using the recently proposed time-dependent quantum Monte Carlo (TDQMC) method which employs sets of particles and quantum waves to describe the ground state and the time evolution of many-electron systems [7-11]. In TDQMC each electron is described statistically as an ensemble of walkers which represent different replicas of that electron in position space, where each walker is guided by a separate time-dependent de Broglie-Bohm pilot wave. The correlated guiding waves obey a set of coupled time-dependent Schrödinger equations (TDSE) where the electron-electron interactions are accounted for using explicit non-local Coulomb potentials. In the TDQMC algorithm the preparation of the ground state of the quantum system involves a few steps which include initialization of the Monte Carlo (MC) ensembles of walkers and guide waves, followed by their concurrent propagation in complex time toward steady state in presence of random component in walker's motion to account for the processes of quantum drift and diffusion. Once the ground state is established, the real-time quantum dynamics can be studied, e.g. the interaction of atoms and molecules with external electromagnetic fields. The large speed up of the calculations when using TDQMC comes from the fact that walker's distribution reproduces the amplitude (or modulus square) of the many-body wave function while its phase is being disregarded as it is not needed for most applications. Also, the TDQMC method can be implemented very efficiently on parallel computers where tens of thousands of coupled Schrödinger equations can be solved concurrently for affordable time.

## 2. General theory

The TDQMC is an *ab initio* method with respect to the electron correlation in that it does not involve explicit pair correlation factors which may become too complex when used for larger systems. For a system of $N$ electrons the many-body wave function obeys the Schrödinger equation:

$$i\hbar \frac{\partial}{\partial t} \Psi(\mathbf{R},t) = -\frac{\hbar^2}{2m} \nabla^2 \Psi(\mathbf{R},t) + V(\mathbf{R})\Psi(\mathbf{R},t) \ , \qquad (2)$$

where $\nabla = (\nabla_1, \nabla_2, ..., \nabla_N)$. The potential $V(\mathbf{R})$ in Eq. (2) is a sum of electron-nuclear, electron-electron, and external potentials:

$$V(\mathbf{r}_1,...,\mathbf{r}_N) = V_{e-n}(\mathbf{r}_1,...,\mathbf{r}_N) + V_{e-e}(\mathbf{r}_1,...,\mathbf{r}_N) + V_{ext}(\mathbf{r}_1,...,\mathbf{r}_N,t)$$

$$= \sum_{k=1}^{N} V_{e-n}(\mathbf{r}_k) + \sum_{k>l}^{N} V_{e-e}(\mathbf{r}_k - \mathbf{r}_l) + V_{ext}(\mathbf{r}_1,...,\mathbf{r}_N,t). \tag{3}$$

For Hamiltonians with no explicit spin variables the exchange effects can be accounted for efficiently using screened Coulomb potentials as described in Ref. 9. The simple idea behind this approach is that the short-range screened Coulomb potential ensures full-scale Coulomb interaction between only electron replicas (MC walkers) which are not too close to each other, in accordance with Pauli's exclusion principle. The use of screened Coulomb potentials is beneficial in that it eliminates the need of using anti-symmetrized products of guiding waves in the Broglie-Bohm guiding equation for the velocity of the walkers. Instead, the many-body wave function is replaced by a simple product:

$$\Psi^k(\mathbf{r}_1,\mathbf{r}_2,...,\mathbf{r}_N,t) = \prod_{i=1}^{N} \varphi_i^k(\mathbf{r}_i,t), \tag{4}$$

where $\varphi_i^k(\mathbf{r}_i,t)$ denote the individual time-dependent guide waves with indexes $i$ and $k$ for the electrons and the walkers, respectively. Then, the guiding equations for the Monte Carlo walkers read:

$$\mathbf{v}(\mathbf{r}_i^k) = \frac{\hbar}{m} \text{Im} \left[ \frac{1}{\varphi_i^k(\mathbf{r}_i,t)} \nabla_i \varphi_i^k(\mathbf{r}_i,t) \right]_{\mathbf{r}_i = \mathbf{r}_i^k(t)}. \tag{5}$$

On the other side, the guide waves obey a set of coupled TDSE:

$$i\hbar\frac{\partial}{\partial t}\varphi_i^k(\mathbf{r}_i,t) = \left[-\frac{\hbar^2}{2m}\nabla_i^2 + V_{e-n}(\mathbf{r}_i) + \sum_{j\neq i}^{N} V_{e-e}^{eff}[\mathbf{r}_i - \mathbf{r}_j^k(t)] + V_{ext}(\mathbf{r}_i,t)\right]\varphi_i^k(\mathbf{r}_i,t), \qquad (6)$$

where the effective electron-electron potential $V_{e-e}^{eff}[\mathbf{r}_i - \mathbf{r}_j^k(t)]$ is expressed as a Monte Carlo sum over the smoothed walker distribution [8]:

$$V_{e-e}^{eff}[\mathbf{r}_i - \mathbf{r}_j^k(t)] = \frac{1}{Z_j^k}\sum_{l=1}^{M} V_{e-e}^{scr}[\mathbf{r}_i - \mathbf{r}_j^l(t)]\, K\!\left(\frac{\left|\mathbf{r}_j^l(t) - \mathbf{r}_j^k(t)\right|}{\sigma_j^k\!\left(\mathbf{r}_j^k,t\right)}\right), \qquad (7)$$

where:

$$Z_j^k = \sum_{l=1}^{M} K\!\left(\frac{\left|\mathbf{r}_j^l(t) - \mathbf{r}_j^k(t)\right|}{\sigma_j^k\!\left(\mathbf{r}_j^k,t\right)}\right), \qquad (8)$$

where $K$ is a smoothing kernel and $Z_j^k$ is the weighting factor. The width $\sigma_j^k\!\left(\mathbf{r}_j^k,t\right)$ of the kernel in Eq. (7) is a measure for the characteristic length of nonlocal quantum correlations within the ensemble of walkers which represent the j-th electron. In practice, the parameter $\sigma_j^k\!\left(\mathbf{r}_j^k,t\right)$ is determined by variationally minimizing the ground state energy of the quantum system [11].

In our calculation a Coulomb potential screened by an error-function is used [9]:

$$V_{e-e}^{scr}[\mathbf{r}_i - \mathbf{r}_j^l(t)] = V_{e-e}[\mathbf{r}_i - \mathbf{r}_j^l(t)] erf\left[\frac{|\mathbf{r}_i - \mathbf{r}_j^l(t)|}{r_j^s \delta_{si,sj}}\right] , \qquad (9)$$

where the Kronecker symbol $\delta_{si,sj}$ restricts the screening effect to the repulsion between only the same-spin walkers, while the value of screening parameter $r_i^s$ is estimated from the Hartree-Fock approximation.

In the approach outlined above, a self-interaction-free dynamics in physical space is achieved where the separate walkers do not share guiding waves which represent different distributions. In order to calculate the many-body probability distribution in configuration space, a separate auxiliary set of walkers with primed coordinates $\mathbf{r}_i'^k$ is introduced which is guided by an anti-symmetric wave function:

$$\mathbf{v}'(\mathbf{r}_i'^k) = \frac{\hbar}{m} \text{Im}\left[\frac{1}{\Psi'^k(\mathbf{r}_1',...,\mathbf{r}_N',t)} \nabla_i \Psi'^k(\mathbf{r}_1',...,\mathbf{r}_N',t)\right]_{\mathbf{r}_j' = \mathbf{r}_j'^k(t)}, \qquad (10)$$

where $\Psi'^k(\mathbf{r}_1',...,\mathbf{r}_N',t)$ is an anti-symmetrized product (Slater determinant or a sum of Slater determinants) of the time-dependent guide waves $\varphi_i^k(\mathbf{r}_i,t)$ of Eq. (6):

$$\Psi'^k(\mathbf{r}_1',\mathbf{r}_2',...,\mathbf{r}_N',t) = A\prod_{i=1}^{N}\varphi_i^k(\mathbf{r}_i',t). \qquad (11)$$

From Eq. (10) and Eq. (11) one can see that each walker with primed coordinates samples the many-body wave function and thus it belongs to all guide waves (i.e. it represents an indistinguishable electron). The distribution of these walkers can be used to directly estimate the average in Eq. (1) by reducing it to (for states with spherical symmetry):

$$I(u,t) \propto \sum_i K_i \left[ \frac{|r_{12}^{\prime i}(t) - u|}{\sigma_{12}^i} \right], \qquad (12)$$

where $r_{12}^{\prime i}(t) = |\mathbf{r}_1^{\prime i}(t) - \mathbf{r}_2^{\prime i}(t)|$. In other words, the pair density function can be simplified to a smoothed histogram (or a kernel density estimation with kernel $K_i$ and bandwidth $\sigma_{12}^i$ [12]) over the ensemble of the distances between the primed walkers.

### 3. Exchange and Coulomb correlation in helium

The two major sources of electron-electron correlation are due to the symmetry of the quantum state and due to the Coulomb repulsion. Here we consider first the effect of the exchange correlation on the pair density function of helium atom. Although the electron pair densities for helium have been analyzed by different techniques they have never, to the author's knowledge, been studied using time-dependent methods.

In order to examine the electron correlation which is due to the exchange interaction we consider the spin-triplet ground state of helium (ortho-helium). The preparation of the ground state is described elsewhere [10]. In the calculation here we use up to 100 000 Monte Carlo walkers and the same number of guiding waves, which are propagated over 2000 complex time steps (Eq. (5) through Eq. (10)) in the presence of random component in walker's motion such that each walker samples the distribution given by its own guiding wave. In order to determine the screening parameter $r_i^s$ of Eq. 9 we invoke the Hartree-Fock approximation where for $\sigma_j^k(\mathbf{r}_j^k,t) \to \infty$ the Coulomb potential in Eq. (7) reduces to a simple (un-weighted) sum of the Coulomb potentials due to all walkers. Because of the spherical symmetry of the $2\,^3S$ state $r_i^s$ is being varied until minimizing the mean integrated squared error of the walker's distribution against the probability distribution obtained from an independent Hartree-Fock solution (e.g. in [13]). Figure 1 shows the probability distributions obtained from TDQMC for the optimizing value of $r_i^s = r^s = 1.13 a.u.$ in Eq. (9). The blue and the green lines show the densities of the walkers guided in physical space (Equations (5) through (9)), respectively, while the red line represents the radial distribution of the walkers guided in configuration space (Eq. (10)). In these calculations a new accurate algorithm for kernel density estimation was used [14]. Notice that all probability distributions throughout this paper are normalized to unity.

The electron pair density for the ground state was calculated very efficiently by simply performing kernel density estimation over the ensemble of distances between the primed walkers. The result is shown in Fig. 2 (a) where the blue and the red lines present the cases with and without exchange interaction, respectively. The lack of exchange ($r_i^s \to 0$ in Eq. (9)) leads to a full (unscreened) Coulomb repulsion, which in the limit of infinite non-local correlation length ($\sigma_j^k(\mathbf{r}_j^k,t) \to \infty$) becomes equivalent to the Hartree approximation. Figure 2 (b) shows the difference between the two curves in Fig. 2 (a), which in fact depicts the shape of the exchange hole for the $2\,^3S$ state of helium (see also e.g. Ref. 4). Note that the exchange hole in our calculation may differ from other results because the distribution of the Monte Carlo walkers varies in radial direction as $r^2R^2(r)$ instead of as $R^2(r)$, where $R(r)$ is the radial wave function. The green line in Fig. 2 (b) shows the exchange hole obtained from an independent Hartree-Fock calculation with no potential screening. It is seen that the two curves are close where the deviations for larger inter-electronic distances are mainly due to the fast decrease of the walker's density away from the core. As the screening parameter $r_i^s$ tends to zero both the height and the width of the exchange hole decrease until the two curves in Fig. 2 (b) become very close, with the only remaining difference being a result of purely Coulomb correlations.

For the ground state of the $2\,^1S$ (para-) helium, the quantity of interest is the Coulomb hole which occurs due to the repulsion of the closely spaced walkers. Figure 3 shows the probability distribution of the ground state walkers as compared to the Hartree-Fock calculation, while Figure 4 (a) depicts the corresponding inter-electronic distances for the two cases. The Coulomb hole calculated as the difference between the two curves is presented in Fig. 4 (b) which is close to previous results by other methods [2]. As the non-local correlation length $\sigma_j^k\left(\mathbf{r}_j^k,t\right)$ tends to infinity both the height and the width of the Coulomb hole decrease until the two curves in Fig. 4 (b) coincide. Thus in our approach where the exchange and the Coulomb correlations are accounted for by solely modifying the potential of electron-electron interaction in physical space, the two parameters $r_i^s$ and $\sigma_j^k\left(\mathbf{r}_j^k,t\right)$ may ensure a smooth transition between the Hartree, the Hartree-Fock, and the fully correlated approximations to the electron-electron interaction. It is important to point out that in the $\hbar/m \to 0$ limit the quantum drift in Eq. (6) vanishes and so does the width of the quantum wave-packet. Therefore, for an isolated atom the quantum correlation length $\sigma_j^k(\mathbf{r}_j^k,t)$ tends to zero in this limit, and if there are no exchange effects ($r_i^s \to 0$), the ensemble of quantum particles governed by Eq. (5) and Eq. (6) transforms to an ensemble of classical particles with the only force being due to the standard Coulomb repulsion between these particles.

## 4. Conclusions

In this paper, it has been shown that for charged particles, the quantum correlation effects which occur due to the exchange and Coulomb correlations can adequately be described by sets of de Broglie-Bohm walkers within the time-dependent quantum Monte Carlo framework. A short-range screening of the Coulomb potential ensures that each replica of a given electron interacts with only those replicas of the rest of the same spin electrons which are sufficiently apart to respect the Pauli's exclusion principle in space. On the other hand, the electron-electron interaction is modified by the quantum non-locality which demands that each replica of a given electron interacts with the replicas of the other electrons which are within the range of the nonlocal quantum correlation length. This concept allows one to build a robust self-consistent and self-interaction-free approach to finding both the ground state and the time evolution of quantum systems. It is demonstrated here that the otherwise awkward procedure for calculating the pair distribution functions of para- and ortho-helium atom can be simplified to the level of finding the ground state probability distributions of the corresponding Monte Carlo walkers.

Besides the relative ease of its implementation, another advantage of using TDQMC is the affordable time scaling it offers which is almost linear with the system dimensionality. This is especially valid when using multicore parallel computers where little communication overhead between the different processes can be achieved, thus utilizing the inherent parallelism of the Monte Carlo methods. This nears the TDQMC to other efficient procedures for treating many-body quantum dynamics such as the time-dependent density functional approximation which, however, suffers systematic self-interaction problems due to the semi-empirical character of the exchange-correlation potentials.

## Acknowledgments

The author gratefully acknowledges support from the National Science Fund of Bulgaria under Grant DCVP 02/1 (SuperCA++). Computational resources from the National Supercomputer center (Sofia) are gratefully appreciated.

**References**

1. A. J. Coleman, Int. J. Quantum Chem. S **1**, 457 (1967). A. J. Thakkar, in *Density Matrices and Density Functionals*, edited by R. Erdahl and V. H. Smith, Jr. (Reidel, New York, 1987), pp. 553–581.

2. C. A. Coulson and A. H. Neilson, Proc. Phys. Soc. London **78**, 831 (1961).

3. R. J. Boyd and C. A. Coulson, J. Phys. B **7**, 1805 (1973).

4. N. Moiseyev, J. Katriel and R. J. Boyd, J. Phy. B **8**, L130 (1975).

5. P. M. W. Gill, D. O'Neill, and N. A. Besley, Theor. Chem. Acc. **109**, 241 (2003).

6. B. M. Austin, D. Yu. Zubarev, and W. A. Lester, Jr., Chem. Rev. **112**, 263 (2012).

7. I. P. Christov, Opt. Express **14**, 6906 (2006).

8. I. P. Christov, J. Chem. Phys. **128**, 244106 (2008).

9. I. P. Christov, J. Phys. Chem. A **113**, 6016 (2009).

10. I. P. Christov, J. Chem. Phys. **135**, 044120 (2011); **135**, 149902 (2011).

11. I. P. Christov, J. Chem. Phys. **136**, 034116 (2012).

12. B. W. Silverman, *Density Estimation for Statistics and Data Analysis,* Monographs on Statistics and Applied Probability (Chapman and Hall, London, 1986).

13. S. E. Koonin and D. C. Meredith, *Computational Physics*, (Addison-Wesley, 1990).

14. Z. I. Botev, J. F. Grotowski and D. P. Kroes, The Annals of Statistics **38**, 2916 (2010).

**Figure captions:**

**Figure 1**. Radial electron density for the ground state of ortho-helium, for MC walkers guided in physical space (blue and green lines) and for MC walkers guided in configuration space (red line). The inset shows the projection of the coordinates of the MC walkers in the x-y plane.

**Figure 2.** Electron pair-density as function of the inter-electronic distance, for the ground state of ortho-helium (a): red line - no screening (no exchange), blue line – short range screened Coulomb potentials. Exchange hole (b) for screened Coulomb potentials (black) and for Hartree-Fock exchange (green).

**Figure 3**. Radial electron density for the ground state of para-helium, for MC walkers guided in physical space (red line) and from the Hartree-Fock approximation (blue line). The inset shows the projection of the coordinates of the MC walkers in the x-y plane.

**Figure 4.** Electron pair-density as function of the inter-electronic distance for the ground state of para-helium (a): red line – correlated result, blue line - Hartree-Fock approximation. The Coulomb hole - (b).

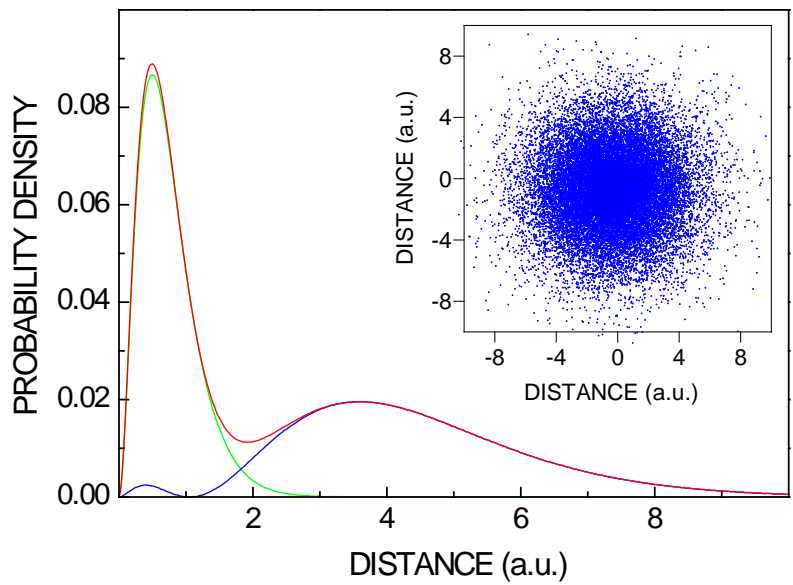

Christov, Figure 1

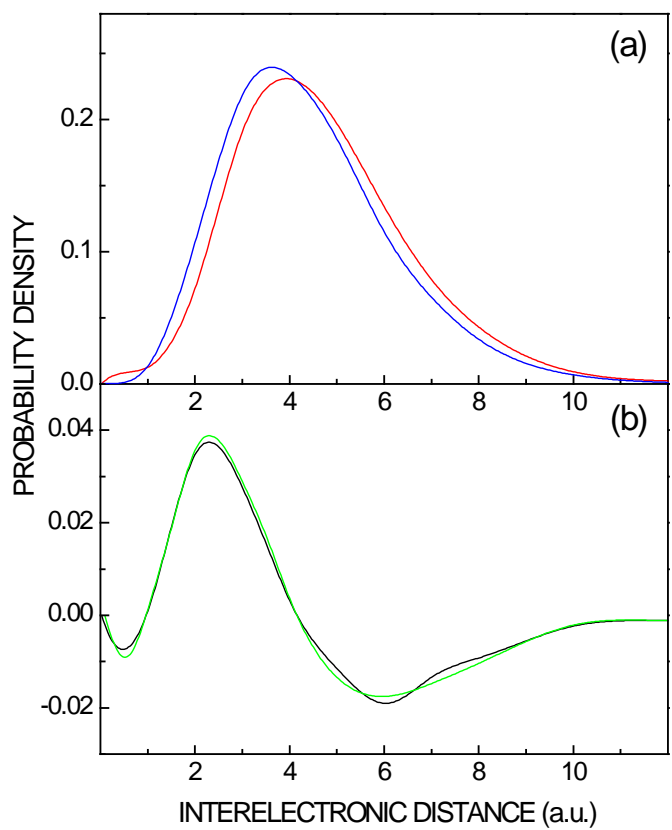

Christov, Figure 2

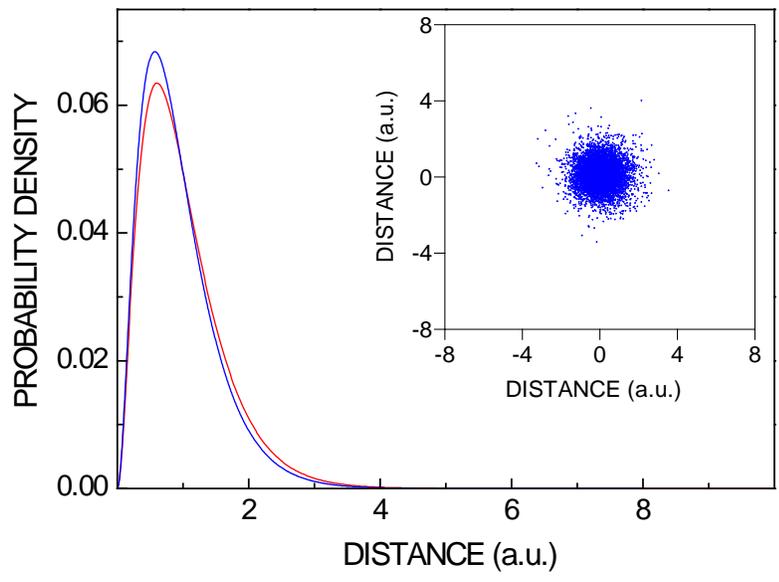

Christov, Figure 3

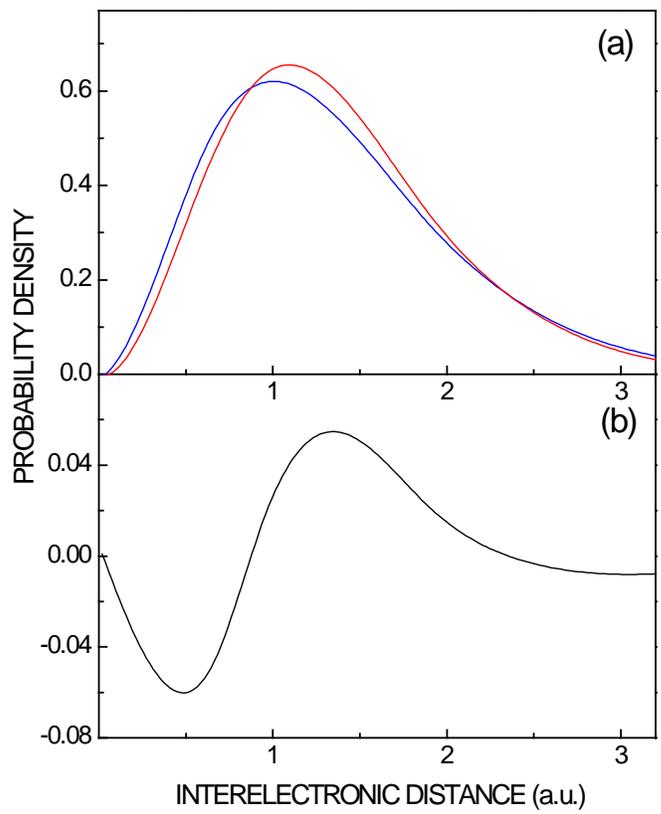

Christov, Figure 4